\documentclass[aps,preprint,amsfonts,amssymb,amsmath]{revtex4}

\pdfoutput=1
\usepackage{color}
\newcommand{\n}{\nonumber}
\newcommand{\be}{\begin{equation}}
\newcommand{\ee}{\end{equation}}
\newcommand{\bea}{\begin{eqnarray}}
\newcommand{\eea}{\end{eqnarray}}

\newcommand{\fD}{~_0D^{1-\alpha}_t}

\begin{document}

\title{ Multi-parameter isospectral Fokker-Planck equations}
\author{Choon-Lin Ho}%\\
%\addresss
\affiliation{Department of Physics, Tamkang University,
Tamsui 25137, Taiwan}

%\date{2025/6/8} 

%\maketitle  % for LaTex

\begin{abstract}

From a given Fokker-Planck equation, a multi-parameter deformed partner Fokker-Planck equation is constructed. 
 This is done by first deleting a set of eigenstates of the original FPE by the multi-step Darboux-Crum transformation, and then reinstating the eigen-energy levels by the reverse Darboux-Crum transformation.  Extension to fractional Fokker-Planck equation is briefly discussed.  A recent study of the one-parameter isospectral FPE applied to black hole in the thermal potential approach is commented.
 
\end{abstract}

% 05.10.Gg  Stochastic analysis methods (Fokker-Planck, Langevin, etc.)
% 02.30.Ik    Integrable systems
% 05.90.+m  Other topics in statistical physics, thermodynamics, and nonlinear dynamical systems (restricted to new topics in section 05)
% 02.50.Ey   Stochastic processes


 \maketitle

%%%%%%%%%%%%%%%%%%%%%%%%%%%%%%%%%%%%

%  Section I --Introduction

%%%%%%%%%%%%%%%%%%%%%%%%%%%%%%%%%%%%

\section{Introduction}

The Fokker-Planck  equation (FPE)  is one of the basic equations used to study various phenomena that involve fluctuations and dissipation
\cite{Ris,Sau}.  As such phenomena are quite common in nature, it is therefore not surprising that FPE  has found applications in such diverse areas as physics, astrophysics, chemistry, biology, finance, etc.  

As with any equation we encounter, it is generally not easy to find analytic solutions of the FPE. Exact analytical solutions of the FPEs are known for only a few cases, such as linear drift and constant diffusion coefficients \cite{Ris,Sau}.

One way to solve the FPE  with time-independent drift and diffusion coefficients is to transform the FPE into a stationary Schr\"odinger equation, and then solve the eigenvalue problem of the latter. Several FPEs have been exactly solved in this way when the associated schr\"odinger equations are solvable \cite{Ris,Sau}. Recently, Such connection has been  employed to investigate quasi-exactly solvable  FPEs \cite{ho1}, and FPEs  with solutions based on the newly discovered exceptional orthogonal polynomials \cite{ho2,ho3}. 

Furthermore, such correspondence between FPE and Schr\"odinger equation allows one to exploit the formalism of supersymmetric (SUSY)  quantum mechanics \cite{SUSY,Sa}. In SUSY quantum mechanics, the  spectrum of a Schr\"odinger equation can be related to that of another through the Darboux transformation \cite{Dar,Cru}.  The SUSY method was first employed in \cite{BB} to compute  the small eigenvalue of the Fokker-Planck equation which controls the rate at which equilibrium is approached. 
Since then there appear  many studies employing SUSY method in various ways to study the FPE, see eg., \cite{K,SMJ,Sch,IN,ho4,ho5} and references therein.
Particularly, solutions of FPEs related by a  one-parameter isospectral  SUSY transformation \cite{Le} were studied in \cite{Ros,R,P}. 

In  Ref. \cite{ho4}  we have made use of the Darboux transformation to find  solutions of a FPE from the solutions of another FPE.
FPEs with either time-independent or dependent drifts (and constant diffusion) were considered. In the case of time-independent drift,  we construct a hierarchy of Darboux transform-related FPEs using the property of shape invariance.  Shape invariance is a sufficient condition for exact solvability, which amazingly is possessed by  most of the exactly-solvable one-dimensional quantum systems. In \cite{ho5} we consider a family of one-parameter deformation from one FPE to its SUSY partner FPE.   
  
In this work we would like to consider relating two FPEs with time-independent drift and constant diffusion by multi-parameter isospectral transformation \cite{Le}, generalizing  the work in \cite{Ros}, which considers a one-parameter transformation.
 By multi-parameter isospectral transformation we mean that the transformed FPE has the same eigenvalues as the original one but with different forms of eigenstates.  This is done by first deleting a set of eigenstates of the original FPE by the multi-step Darboux tranformation, which is usually called the Darboux-Crum transform, and then reinstating the eigen-energy levels by the reverse Darboux transformation.  
 
We will briefly review the connection between FPE and Schr\"odinger equation in Sect.\,II.  Sect.\,III explains the main ideas of  construction of FPE pair based on the Darboux-Crum transformation.  Multi-parameter Isopectral FPEs are discussed in Sect.\,IV.  These procedures are then extended to a class of fractional FPE in Sect.\,V. In the last section we summarize the paper, and briefly comment on a recent attempt to apply the isopectral transformation to black holes in thermal potential. 

%%%%%%%%%%%%%%%%%%%%%%%%%%%
\section{FPE and Schr\"odinger equation}

The FPE with constant diffusion coefficient (which we set to be unity without loss of generality) and  time-independent drift $D^{(0)}(x)$ is
\be
\frac{\partial}{\partial t} P(x,t)=\left[-\frac{\partial}{\partial x} D^{(0)}(x) +
\frac{\partial^2}{\partial x^2}\right]P(x,t).
\label{FPE}
\ee
The domain of $x$ is taken to be $x\in [c_1, c_2]$.  Typical values of $c_1, c_2$ are $c_1=0, -\infty$ and $c_2>0, +\infty$.
It is understood that, as probability density function (PDF),  $P(x,t)$ is to be normalized over the domain of $x$.

It is convenient to write  $D^{(0)}(x)=-2W^\prime (x)\equiv -U^\prime(x)$,    where the prime denotes $x$-derivative.   The function $U(x)\equiv 2W(x)$ is called the drift potential of the system. Putting
\be
P(x, t)=e^{-W(x)}\exp(-\epsilon t)\phi(x)
\label{P}
\ee
in (\ref{FPE}), 
we obtain
\bea
&& ~~~~~ H\phi =\epsilon\phi,\label{H}\\
H &\equiv& -\partial_x^2+W^{\prime 2} - W^{\prime\prime}.\n
\eea
For clarity of presentation, we often left out arguments in formulas in what follows.

The function $\phi$ satisfies the stationary Schr\"odinger equation with Hamiltonian $H$ and eigenvalue $\epsilon$.
The Hamiltonian $H$ can be written as
$H=A^{(0)\dagger} A^{(0)}$, with the pair of Hermitian conjugate operators
\be
 A^{(0)}=\partial_x  +W^\prime, ~~A^{(0)\dagger}=-\partial_x  + W^\prime.
 \ee
Taking the overlap integral of $\phi$ with the first equation in (\ref{H}),  we have 
\be
\epsilon \langle\phi |\phi\rangle=\langle \phi | H\phi\rangle=\langle \phi |A^{(0)\dagger} A^{(0)}\phi\rangle
= \langle A^{(0)}\phi | A^{(0)}\phi\rangle \geq 0.
\ee
Since $\langle\phi | \phi\rangle >0$ for non-vanishing states, the above relation then implies that the eigenvalues are non-negative, i.e. $\epsilon \geq 0$.  We will arrange $\epsilon$ in increasing order : $0\leq \epsilon_0< \epsilon_1<\epsilon_2<\ldots$.

The function $\phi_0\equiv \exp(-W)$ is the zero
mode of $H$: $H\phi_0=0$.  If $\phi_0$ is non-normalizable, then it is not an eigenstate of $H$, and $\epsilon_0=0$ is absent in the spectrum.  For definiteness, we consider only the case when $\phi_0$ is an eigenstate.  In this case, $W=-\ln|\phi_0|$, and 
\be
 A^{(0)}=\partial_x  -(\ln|\phi_0|)^\prime, ~~A^{(0)\dagger}=-\partial_x  -(\ln|\phi_0|)^\prime.
 \ee
 Note that $A^{(0)}\phi_0=0$.  The drift is related to the ground state by  $D^{(0)}(x)=2(\ln |\phi_0|)^\prime$.
 We mention here that the construction to be discussed below can be easily extended to the case where $\epsilon_0=0$ is absent
with slight modification using $\phi_1$ in place of $\phi_0$.

This connection of FPE with the Schr\"odinger equation allows the FPE be solved by the method of eigenfunction expansion. If all the eigenfunctions $\phi_k$
($k=0,1,2,\ldots$) of $H$ with eigenvalues $\epsilon_k$ are solved, then the solution of  $P_k(x,t)$  of (\ref{FPE}) corresponding to
the eigenvalue $\epsilon_k$ is $P_k(x,t)=\phi_0(x)\phi_k(x)\exp(-\epsilon_k t)$.  The stationary distribution is $P_0(x)=\phi_0^2=\exp(-2W)$, which is
obviously non-negative. Any positive definite initial probability density $P(x,0)$ can be expanded as $P(x,0)=\phi_0(x)\sum_k c_k\phi_k(x)$, with constant coefficients $c_k$ ($k=0,1,\ldots$)
\begin{eqnarray}
c_k=\int_{-\infty}^\infty \phi_k(x)\left(\phi_0^{-1}(x)
P(x,0)\right)dx.
\label{c}
\end{eqnarray}
Then at any later time $t$, the solution of the FP equation is
\be
P(x,t)=\phi_0(x)\sum_k c_k \phi_k(x)\exp(-\epsilon_k t).
\label{sol}
\ee

In this work we consider how to obtain solutions of a FPE from those of another by the Darboux-Crum and isospectral transformations between the two Schr\"odinger equations associated with them.

%%%%%%%%%%%%%%%%%%%%%%%%%
\section{Darboux-Crum transformation}

In \cite{Sch} the construction of Darboux-Crum pair of FPEs is considered by  deleting a set of auxiliary states of the Schr\"odinger equation.  Since our purpose here is the isospectral FPEs, we shall consider only Darboux-Crum transformation using the lowest lying eigenstates.  To make our presentation self-contained, and to fix notations, we shall briefly review this transformation based on the set of lowest lying states \cite{Cru,Sa}.

As explained in the last section, the Hamiltonian $H$ can be written as
$H=A^{(0)\dagger} A^{(0)}$. The Darboux theorem states that the Hamiltonian $H^{(1)}=A^{(0)} A^{(0)\dagger} $  has (unnormalized) eigenstates
 \be
 \phi_k^{(1)} = A^{(0)}\phi_k,  ~~ A^{(0)}\phi_0 = 0,    ~~~k=1,2,\ldots,
 \ee
  with eigenvalues $\{\epsilon_1, \epsilon_2,\ldots\}$. Thus $H^{(1)}$ 
 has the same spectrum as $H$ except $\epsilon_0=0$.  $H^{(1)}$ is called the  Darboux pair of $H$ obtained by deleting the ground state $\phi_0$ of $H$.
 
 Note that $H^{(1)}$ can also be expressed as $H^{(1)}=A^{(1)\dagger} A^{(1)\dagger} +\epsilon_1$, where  
\be
 A^{(1)}=\partial_x  -(\ln|\phi^{(1)}_1|)^\prime, ~~A^{(0)\dagger}=-\partial_x  -(\ln|\phi^{(1)}_1|)^\prime.
 \ee
We have $A^{(1)}\phi^{(1)}_1=0$.
To construct a FPE based on $H^{(1)}$ but having a ground state at zero energy, we subtract $\epsilon_1$ from $H^{(1)}$ and  take as  Hamiltonian $H_1=A^{(1)\dagger} A^{(1)}$ , which has spectrum $\{\epsilon_k-\epsilon_1, k=1,2,\ldots\}$.  This Hamiltonian then corresponding to a FPE with a drift coefficient $D^{(1)}=2(\ln|\phi_1^{(1)}|)^\prime$.

Similarly, by deleting the ground state $\phi^{(1)}_1$ of $H_1$, one can obtain a new Hamiltonian $H_2=A^{(2)\dagger} A^{(2)}$ with
\bea
A^{(2)}&=&\partial_x  -(\ln|\phi^{(1)}_1|)^\prime,\n\\
\phi^{(2)}_k&=&A^{(1)}\phi^{(1)}_k, ~~k=2,3,\ldots,\\
&&A^{(2)}\phi^{(2)}_2=0,\n
\eea
and spectrum $\{\epsilon_k-\epsilon_2, k=2,3,\dots\}$.  This Hamiltonian corresponds to a FPE with $D^{(2)}=2(\ln|\phi^{(2)}_2|)^\prime$.

This process can be continued by successively deleting the lowest lying energy states of the original hamiltonian $H$.    Suppose the lowest $n$ states $\phi_s, s=0,1,\ldots, n-1$ are deleted, then the $n$-step Hamiltonian $H_n=A^{(n)\dagger} A^{(n)}$ will have eigenstates and eigenvalues given by
\bea
\phi^{(n)}_k&=&A^{(n-1)}\phi^{(n-1)}_k, ~~k=n, n+1, \ldots,\n\\
A^{(n)}&=&\partial_x  -(\ln|\phi^{(n)}_n|)^\prime,\\
&&A^{(2)}\phi^{(2)}_2=0,\n
\eea
and spectrum $\{\epsilon_k-\epsilon_n, k=n, n+1,\dots\}$.
Now it is the interesting result by Crum that the  states $\phi^{(n)}_k$ of $H_n$ can be directly related to the states of $H$ as
\cite{Cru}
\bea
\phi^{(n)}_k&=& A^{(n-1)}A^{(n-2)}\cdots A^{(1)}A^{(0)}\phi_k\n\\
&=&\frac{W\left[\phi_0,\phi_1, \ldots, \phi_{n-1}, \phi_k\right](x)}{W\left[\phi_0,\phi_1, \ldots, \phi_{n-1}\right](x)},\label{crum}\\
A^{(s)}&=&\partial_x  -(\ln|\phi^{(s)}_s|)^\prime, ~~s=0,1,\ldots,n-1.\n
\eea
Here $W\left[f_1, f_2, \ldots, f_m\right](x)$ is the Wronskian of the the functions $f_1, f_2, \ldots, f_m$.

We will call the FPEs corresponding to $H$ and $H_n$ the $n$-step Darboux-Crum pair.
The drift coefficient of the partner FPE is $D^{(n)}=2(\ln|\phi_n^{(n)}|)^\prime$.
Given  any solution $P(x,t)$ (\ref{sol}) of the original FPE (\ref{FPE}), we can associate a corresponding PDF of the partner  FPE by 
\bea
P^{(n)}(x,t) &\sim& \phi^{(n)}_n  \sum_{k\geq n} c_k \phi^{(n)}_k(x)\exp[-(\epsilon_k- \epsilon_n )t]\label{P_DC}\\
   &=& e^{\epsilon_n t} \phi^{(n)}_n A^{(n-1)}A^{(n-2)}\cdots A^{(1)}A^{(0)} \left(\phi_0^{-1} P(x,t)\right).\n
 \eea
Note the first $n$ terms in $P(x,t)$ are annihilated by the operators $A^{(s)} (s=0,1,\ldots, n-1)$.

%%%%%%%%%%%%%%%%%%%%
\section{Isospectral pair of FPEs}
\label{IsoFPE}

We now come to a different way to associate two FPEs. In the Darboux-Crum transformation, a set of lowest lying states in $H$ are deleted to obtain $H_n$.  However, it is possible to reinstate these energy levels and obtain a Hamiltonian that has exactly the same spectrum as that of $H$, but with different eigenstates.  We shall call this the isospectral transfroamtion.
One-parameter isospectral FPEs have been studied in \cite{Ros,R,P}.  Here we construct FPE and its solutions which are multi-parameter isospectral transforms of those of a given FPE.

Construction of isospectral Hamiltonians has been well explained in \cite{Le,SUSY} . The main concern there is  the derivation of the new ground state, and the Hamiltonian constructed from it. Other eigenstates are not given. However,  in our case, we need to know also all these other eigenstates for the solutions of FPE as in (\ref{sol}).  So here we will recapitulate the essence of the multi-parameter isospectral transform, with expressions of all eigenstates explicitly given.

%-------------------------
\subsection{$1$-parameter isospectral FPE pair}
\label{1-para}

From the previous section we see that
$H^{(1)}=A^{(0)}A^{(0)\dagger}  =A^{(1)\dagger} A^{(1)} +\epsilon_1$ has the same spectrum as 
$H=A^{(0)\dagger} A^{(0)}$, except the ground state $\phi_0$ with energy  $\epsilon_0=0$. 
However, from $A^{(0)\dagger} \phi^{-1}_0=0$, it  is seen that  $\phi^{-1}_0$ is an eigenstate of $H^{(1)}$ with 
eigenvalue $\epsilon_0=0$, though it is not normalizable. 
In fact, the most general solution satisfying $A^{(0)\dagger} \Phi(\lambda_0)=0$ is
\be
\Phi_0(x; \lambda_0)= (I_0+\lambda_0)/\phi_0, 
\ee
where 
\be
I_0(x)\equiv \int^x_{c_1} dx^\prime\,\phi_0^2(x^\prime),
\ee
$\lambda_0$ is a real parameter, and $c=0$ or $-\infty$ for the half or full line, respectively.

We can regard $\Phi_0(\lambda_0)$ as a virtual eigenstate of $H^{(1)}$ corresponding to $\epsilon_0=0$.  Note that $\Phi^{-1}_0(\lambda_0)$ is a solution of $B_0(\lambda_0)\Phi^{-1}_0(\lambda_0)=0$, where
\be
B_0(\lambda_0)=\partial_x  -(\ln|\Phi^{-1}_0(\lambda_0)|)^\prime,
\ee
and  is normalizable for $\lambda_0 \notin [-I_0(c_2),0]$.

So we can  consider $\Phi^{(-1)}_0(\lambda_0)$ as the ground state of a Hamiltonian $\hat{H}^{(0)}\equiv B^\dagger_0(\lambda_0)B_0(\lambda_0)$.     By applying the inverse Darboux transformation on $H^{(1)}$ using $\Phi^{-1}_0(\lambda_0)$ as the base function, it is shown that $\hat{H}^{(0)}$ is indeed a Hamiltonian having the same spectrum as $H$ with eigenstates 
\bea
\hat{\phi}_0 &=& \Phi^{-1}_0(\lambda_0),\n\\
\hat{\phi}_k &=& B^\dagger_0(\lambda_0) \phi^{(1)}_k\\
&=& B^\dagger_0 A^{0)}\phi_k, ~~k=1,2,\ldots.\n
\eea
As before, for simplicity we often suppress arguments in formulas.
 It is usually said in the literature that $\hat{H}^{(0)}$ is the Hamiltonian with a ground state at energy $\epsilon_0=0$ reinstated.

 %------------------------------
 \subsection{$2$-parameter isospectral FPE pair}
 \label{2-para}
 
 To construct a $2$-parameter isospectral partner of $H$, we start from $H^{(2)}=A^{(1)}A^{(1)\dagger}  =A^{(2)\dagger} A^{(2)} +\epsilon_2$ .
 By the procedure in the previous subsection, we first construct a Hamiltonian $\hat{H}^{(1)}$ with an eigenstate corresponding to $\epsilon_1$ reinstated from the virtual state 
\be
  \Phi_1(\lambda_1)= (I_1+\lambda_1)/\phi^{(1)}_1, ~~~I_1(x)\equiv \int^x_{c_1} dx^\prime\,[\phi^{(1)}_1]^2(x^\prime).
\ee
 The results are
 \bea
  \hat{H}^{(1)} &\equiv& B^\dagger_1(\lambda_1)B_1(\lambda_1),\n\\
 B_1(\lambda_1) &=& \partial_x  -(\ln|\Phi^{-1}_1(\lambda_1)|)^\prime, \n\\
\hat{\phi}^{(1)}_1 &=& \Phi^{-1}_1(\lambda_1),\n\\
\hat{\phi}^{(1)}_k &=& B^\dagger_1(\lambda_1) \phi^{(2)}_k\\
&=& B^\dagger_1 A^{(1)}A^{(0)}\phi_k, ~~k=2,3\ldots.\n
 \eea
 
 Now we construct the isospectral hamiltonian $\hat{H}^{(0)}$ by reinstating a state corresponding to $\epsilon_0$. This is done from the inverse Darboux transform of the corresponding virtual state $\hat{\phi}^{1}_0$ of $\hat{H}^{(1)}$.  The virtual state $\hat{\phi}^{1}_0=B^\dagger_1\phi^{(2)}_0$ is the inverse Darboux transform of the virtual state $ \phi^{(2)}_0=A^{(1)}\Phi_0$,  which in turn is the Darboux transform of the virtual state 
 $\Phi_0(\lambda_0)$ of $H^{(1)}$ as given in the previous subsection.  Thus the virtual state of $\hat{H}^{(0)}$ corresponding to $\epsilon_0$ 
 is
 \be
\Phi_0(\lambda_0, \lambda_1)= B^\dagger_1(\lambda_1) A^{(1)} \Phi_0(\lambda_0).
\ee

Applying the inverse Darboux  transformation on $\hat{H}^{(1)}$,  we get the $2$-parameter isospectral Hamiltonian
 \bea
  \hat{H}^{(0)} &\equiv& B^\dagger_0(\lambda_0, \lambda_1)B_0(\lambda_0, \lambda_1),~~~\lambda_{0,1} \notin [-I_{0,1}(c_2),0],\n\\
 B_0(\lambda_0, \lambda_1) &=& \partial_x  -(\ln|\Phi^{-1}_0(\lambda_0, \lambda_1)|)^\prime, \n\\
\hat{\phi}^{(0)}_0&=& \Phi^{-1}_0(\lambda_0, \lambda_1),\\
\hat{\phi}^{(0)}_1 &=& B^\dagger_0(\lambda_0, \lambda_1) \hat{\phi}^{(1)}_1\n \\
                            &=& B^\dagger_0  \Phi^{-1}_1,\n\\         
\hat{\phi}^{(0)}_k &=& B^\dagger_0(\lambda_0, \lambda_1) \hat{\phi}^{(1)}_k\n \\
                            &=& B^\dagger_0 B^\dagger_1 A^{(1)}A^{(0)}\phi_k, ~~k=2,3\ldots\n
 \eea

%------------------------------------------
 \subsection{$n$-parameter isospectral FPE pair}
 \label{n-para}
 
 It is now easy to see how the procedures described in the previous subsections can be extended to multi-parameter cases
 by deleting and reinstating  $n$  lowest lying energy states of $H$, $\{\phi_s, s=0,1,\ldots, n-1\}$. 
 The main data of the transformation are listed below, particularly the explicit expressions of eigenstates not presented in \cite{Le,SUSY}.
 
 Formulas for some basic functions and operators:
  \bea
  \Phi_s(\lambda_s) &=&  (I_s+\lambda_s)/\phi^{(s)}_s, ~ I_s(x)\equiv \int^x_{c_1} dx^\prime\,[\phi^{(s)}_s]^2(x^\prime), ~~~~~ \lambda_s \notin [-I_s(c_2),0],\n\\
   B_s(\lambda_s,\ldots, \lambda_{n-1}) &=& \partial_x  -(\ln|\Phi^{-1}_s(\lambda_s,\ldots, \lambda_{n-1})|)^\prime, 
\n\\
  \Phi_s(\lambda_s,\ldots, \lambda_{n-1}) 
                    &=& B^\dagger_{s+1}(\lambda_{s+1}, \ldots,\lambda_{n-1})  
                    B^\dagger_{s+2}(\lambda_{s+2}, \ldots,\lambda_{n-1}) \cdots B^\dagger_{n-1}(\lambda_{n-1})\\
                    && \times A^{(n-1)}A^{(n-1)}\cdots A^{(s+1)} \Phi_s(\lambda_s),\n
 \eea
 
The isospectral Hamiltonian is then given by
 \be
   \hat{H}^{(0)} \equiv B^\dagger_0(\lambda_0,\ldots, \lambda_{n-1}) B_0(\lambda_0,\ldots, \lambda_{n-1}),
 \ee
 and its eigenstates are
\bea
\hat{\phi}^{(0)}_0   &=& \Phi^{-1}_0(\lambda_0,\ldots, \lambda_{n-1}),\n\\
\hat{\phi}^{(0)}_s   &=& B^\dagger_0 B^\dagger_1\cdots B^\dagger_{s-2}B^\dagger_{s-1} \Phi^{-1}_s(\lambda_s,\ldots, \lambda_{n-1}) ,~~~~s=1,2,\ldots,n-1,\label{n-iso}\\         
\hat{\phi}^{(0)}_k &=& B^\dagger_0 B^\dagger_1\cdots B^\dagger_{n-2}B^\dagger_{n-1} \phi^{(n)}_k,~~~~k\geq n.\n
\eea
Again, for simplicity and clarity, we have suppressed arguments in the above formulas.
By (\ref{crum}), the last expression in (\ref{n-iso}) can be written as
\bea
 && ~~~~~~~~~~\hat{\phi}^{(0)}_k = \mathcal{L}_n \phi_k,~~~~~k\geq n,\label{An}\\     
     \mathcal{L}_n &\equiv&   B^\dagger_0 B^\dagger_1\cdots B^\dagger_{n-2}B^\dagger_{n-1}A^{(n-1)}\cdots A^{(1)}A^{(0)}.\n
\eea

The  FPE defined by the drift coefficient 
$D^{(n)}=2(\ln|\Phi^{-1}_0(\lambda_0,\ldots, \lambda_{n-1})|)^\prime$
is the $n$-parameter isospectral pair of the original FPE (\ref{FPE}).
For any solution $P(x,t)$ (\ref{sol}) of (\ref{FPE}), there corresponds a solution of the isospectral FPE, given by 
\be
P^{(n)}(x,t)\sim  \Phi_0 \left[ \sum_{s=0}^{n-1} c_s  \hat{\phi}^{(0)}_s \exp(-\epsilon_s t)  +  \mathcal{L}_n \left(\phi_0^{-1}P(x,t)\right)\right].
\label{P_Iso}
\ee

 Note that the sum of terms involving the $n$ lowest lying states is listed separately in the square-bracket.  This is because
 they are absent in the second term as the operator ${\cal L}_n$ annihilates them. 

%%%%%%%%%%%%%%%%%%

\section{Extension to a class of Fractional FPE}

In recent years anomalous diffusions have attracted much interest owing to their ubiquitous appearances in many physical situations.
 For examples, charge carrier transport in amorphous semiconductors, nuclear magnetic resonance diffusometry in percolative, and porous systems, Rouse or reptation dynamics in polymeric systems, transport on fractal geometries, and many others \cite{FP0}.
Unlike the  well-known Brownian motion, anomalous diffusions are  characterized by a mean-squared displacement relation   $\langle x^2 \rangle\sim t^\gamma(\gamma\neq 1)$ which is not linear in time --  it is  superdiffusive for $\gamma>1$, and subdiffusive for $\gamma<1$.

One approach to explain the anomalous diffusion is to consider fractional FPE, i.e., FPE with fractional derivatives \cite{FP0,FP1,FP2,FP3,FP4,FP5}.   The fractional FPE proposed in \cite{FP1,FP2} is
\bea
\frac{\partial}{\partial t} P(x,t)&=&\fD \left[-\frac{\partial}{\partial x} D^{(0)}(x)+
\frac{\partial^2}{\partial x^2}\right]P(x,t),\n\\
&& ~~0<\alpha<1.
 \label{FFP}
\eea
Here 
$\fD$ is the Riemann-Liouville fractional derivative defined by \cite{OS}
\be
\fD f(x,t) =\frac{1}{\Gamma(\alpha)}\frac{\partial}{\partial t} \int^t_0 (t-t^\prime)^{\alpha-1} f(x,t^\prime)\, dt^\prime\n
\label{RL}
\ee
for $0<\alpha<1$.

In place of (\ref{P}), we let
\be
P(x, t)=e^{-W(x)}T(t)\phi(x).
\label{fP}
\ee
Then from (\ref{FFP})  we find that $\phi(x)$ satsifes the same equation (\ref{H}), while the temporal part $T(t)$ is determined by
\be
\frac{d}{dt}\,T(t)=- \epsilon \fD T(t).\label{T}
\ee
This  is solved by \cite{FP1}
\be
T(t)=E_\alpha (-\epsilon \,t^\alpha).
\ee
Here $E_\alpha (z)$ is the Mittag-Leffler function
\begin{equation}
E_\alpha(z)=\sum_{k=0}^\infty \frac{z^k}{\Gamma(\alpha k +1)}, ~~\alpha\in \mathbb{C}, ~~{\rm Re}\  \alpha >0,
\end{equation}
which is a one-parameter deformation of the Euler exponential function.

One see that the eigen-solutions of the fractional FPE (\ref{FFP}) are similar to those of the ordinary FPE in (\ref{FPE}), except 
in the temporal factor the Mittag-Leffler function replaces the Euler exponential function.  The Schr\"odinger equation associated with these two types of FPEs are exactly the same. Thus the procedures presented in previous sections can be directly carried over to the fractional FPEs. 
The general solution of Eq.\,(\ref{FFP}) is
\be
P(x,t)=\phi_0(x)\sum_k c_k \phi_k(x) E_\alpha (-\epsilon_k\,t^\alpha)
\label{fP}
\ee
with constant coefficients $c_k$ ($k=0,1,\ldots$) given by (\ref{c}).

The Darboux-Crum FPE pair has drift $D^{(n)}=2(\ln|\phi_n^{(n)}|)^\prime$. As in the ordinary FPE cases,  for each solution (\ref{fP}) one can associated the following solution to the partner fractional FPE,
\bea
P^{(n)}(x,t) &\sim& \phi^{(n)}_n  \sum_{k\geq n} c_k \phi^{(n)}_k(x) E_\alpha [-(\epsilon_k- \epsilon_n )t].
\eea
We note that the second expression in (\ref{P_DC}), with  $A^{(k)}$'s acting on $P(x,t)$, cannot be used here, as that relation relies on the factorization property of the Euler exponential function, namely,  $\exp(x+y)=\exp(x) \exp(y)$, which the Mittag-Leffler function  lacks.

Isospectral fractional FPEs can also be constructed as in Sect.\,\ref{IsoFPE}.  The drift of the partner system is again 
$D^{(n)}=2(\ln|\Phi^{-1}_0(\lambda_0,\ldots, \lambda_{n-1})|)^\prime$, and the solution corresponding to (\ref{fP}) is
\be
P^{(n)}(x,t)\sim  \Phi_0 \left[ \sum_{s=0}^{n-1} c_s  \hat{\phi}^{(0)}_s E_\alpha (-\epsilon_s t)  +  \mathcal{L}_n \left(\phi_0^{-1}P(x,t)\right)\right].
\ee

%%%%%%%%%%%%%%%

\section{Summary and discussion}

By exploiting the connection between FPE and Schr\"odinger equations, we have shown how to relate two FPEs with time-independent drift and constant diffusion by multi-parameter isospectral transformation.  This refer to two FPEs the Schr\"odinger of which  have the same eigenvalues but with different forms of eigenstates.  This is done by first deleting the lowest lying eigenstates set  of the original FPE by the multi-step Darboux-Crum tranformation, and then reinstating the eigen-energy levels by the reverse Darboux-Crum transformation.  
This connection allows us to obtain  a solution of the transformed FPE, once a solution of the original FPE is known.

In this work we have considered Darboux-Crum transformation by deleting and reinstating only the lowest lying levels of the Schr\"odinger equation corresponding to the original FPE.  It should be mentioned that  the Darboux-Crum transformation is not restricted by this requirement -- any set of solutions of the Sch\"odinger equation can be employed for that purpose.   However, our construction can be easily extended to this general case.

While this manuscript is in preparation, a recent paper \cite{SS} came to our attention.  In this work, the one-parameter isospectral transformation is applied to a type of FPE recently proposed to study some dynamic behaviors of black holes \cite{Xu}.  
It is nice to see that isospectral FPEs may play a role in black hole physics, and  it is not difficult to extend the work in \cite{SS} to multi-parameter cases by the procedures described here.  However, we think there are some issues that  deserve clarification and further  study in regard to extending black hole thermal potential by isospectral transformation. 
Our main concern is whether the transformed FPE correspond to real black hole systems,  as explained below.

The main idea of \cite{Xu} is to put an ensemble of thermodynamical states in a thermal potential $U$, and then study the stochastic behavior of the system using a FPE with $U$  as the drift potential. 
One or a group of these states can represent the real black holes systems, which are solutions of the Einstein equation. 
The thermal potential $U$ is defined in \cite{Xu} by
\be
U=\int(T_h-T)dS,
\label{U}
\ee
where $T$ is the temperature of the ensemble treated as an independent parameter, $T_h$ and  $S$ are the Hawking temperature and  the thermodynamic entropy of the real black hole, respectively.   For black holes $T_h=T_h(S,Y)$ is a function of $S$ and other parameters $Y$ of the black hole, such as charge, angular momentum, etc. When $T=T_h$, the ensemble is made up of real black hole states, and the system is in thermal equilibrium. As such, $U$ can be regarded 
as a measure of the deviation of all the states in the ensemble from the real black hole state. For the Schwarzschild black hole, in terms of the radius of its event horizon $r_h$, we have
\be
T_h=\frac{1}{4\pi r_h}, ~~ S=\pi r^2_h,
\ee
giving
\be
U=\frac12 r_h - \pi T r^2_h.
\ee
It appears that the author in \cite{Xu}  considers a FPE in the $r_h$-space with $U$ plays in some way the role of the drift potential, and then uses it to study the dynamic behaviors of the black hole system.   Now in \cite{SS} the one-parameter  isospectral Schr\"odinger equation of the Schr\"odinger equation corresponding to this (Schwarzschild) FPE is  constructed.  The main part of this work is to study the potential and the eigenstates of the isospectral Schr\"odinger equation, as discussed in \cite{Le,SUSY}.  The isospectral FPE was not presented and analyzed. 

What concerns us is that the transformed thermal potential obtained in this way $\hat{U}$ (be it a one- or multi-parameter version) is related to the drift potential of the new FPE the work \cite{SS} intended to construct. 
In most  cases, a drift potential defines a stochastic system.  But in the thermal potential approach to black holes proposed in \cite{Xu},  the new thermal potential $\hat U$ should be linked to a black hole system.  That is, by (\ref{U}) this new FPE should be related to a new black hole system associated with a Hawking temperature $\hat{T}=\hat{T}(\hat{S}, \hat{Y})$. 
It is not clear to us if there is black hole solution to the Einstein equation that satisfies this new relation of $\hat T$. Furthermore, the quantities $\hat{T}_h, \hat{S}, \hat{Y}$ will be functions of the parameters $\{\lambda_0, \lambda_1, \ldots,\lambda_{n-1}\}$. What are the physical meanings of these parameters in the new black hole system?  It seems that further study in this thernal potential approach to black holes would be helpful  to clarify these issues make isopectral transformation useful in black hole physics.

%%%%%%%%%%%%%%%%%%%%
\section*{Acknowledgments}

The work is supported in part by the Ministry of Science and Technology (MOST)
of the Republic of China under Grant NSTC 113-2112-M-032-010.

%%%%%%%%%%%%%%%%%%

%%%%%%%%%%%%%

%-----------------------------
\end{document}